\documentclass[floatfix]{amsart}
\usepackage{graphicx}
\usepackage{amssymb,amsthm}

\title{AC Magnetotransport in Reentrant Insulating Phases of Two-dimensional Electrons near 1/5 and 1/3 Landau fillings}

\author{Yong P. Chen$^{1,2,*}$, Z. H. Wang$^{3,2}$, R. M. Lewis$^{2,1}$, P. D. Ye$^{4}$, L. W. Engel$^{2}$, \\D. C. Tsui$^{1}$, L. N. Pfeiffer$^{5}$ and K. W. West$^{5}$\\}
\address{
$^{1}$Department of Electrical Engineering, Princeton University, Princeton, NJ 08544, USA\\
$^{2}$National High Magnetic Field Laboratory, Tallahassee, FL 32310, USA\\
$^{3}$Department of Physics, Princeton University, Princeton, NJ 08544, USA\\
$^{4}$Agere Systems, Allentown, PA 18109, USA\\
$^{5}$Bell Laboratories, Murray Hill, NJ 07974, USA\\
$^{*}$yongchen@princeton.edu 
}
\begin{document}

\begin{abstract}
We have measured high frequency magnetotransport of a high quality two-dimensional electron system (2DES) near the reentrant insulating phase (RIP) at Landau fillings ($\nu$) between 1/5 and 2/9. The magneto\textit{conductivity} in the RIP has 
resonant behavior around 150 MHz, showing a \textit{peak} at $\nu$$\sim$0.21. 
Our data support the interpretation of the RIP as due to some pinned electron solid.
We have also investigated a narrowly confined 2DES recently found to have a RIP 
at 1/3$<$$\nu$$<$1/2 and we have revealed features, not seen in DC transport, that suggest some intriguing interplay between the 1/3 FQHE and RIP. 
\\
\textbf{keywords:}Reentrant insulating phase; 2DES; high-frequency transport
\end{abstract}

\maketitle

The DC resistivity of high quality two dimensional electron systems (2DES) in GaAs/AlGaAs is well known to diverge at low temperature ($T$) for Landau filling 
$\nu$=$nh/eB$ (where $n$ denotes 2DES density and 
$B$ the perpendicular magnetic field) below\cite{willett} the 1/5 
fractional quantum Hall effect (FQHE) and in a narrow $\nu$ range reentrant above\cite{jiang} 1/5. 
 Such a reentrant insulating phase (RIP) has been interpreted as a pinned electron solid with Wigner crystal (WC)\cite{rev} order, and the reentrance of the insulating transition is thought to be caused by competition between FQHE states and WC\cite{cusp}. Recently it was shown that such a transition can shift to higher $\nu$, and occur around the 1/3 FQHE, for a 2DES tightly confined in a narrow quantum well (QW)\cite{kang}. 

In this article, we report AC magnetotransport measurements on two GaAs/AlGaAs 2DES samples (sample 1 and 2) in the RIP around $\nu$=1/5 and $\nu$=1/3 respectively. 
Sample 1 is a 65-nm-wide QW with $n$=5.9$\times$10$^{10}$cm$^{-2}$ 
and mobility $\mu$$\approx$8$\times$10$^6$cm$^2$/Vs. Samples$^{2,6-8}$ 
with such high quality ($\mu$ over 10$^6$cm$^2$/Vs) have been well known to display (in DC) a RIP at $\nu$$>$1/5.
Sample 2 is an 8-nm-wide QW with $n$=1.2$\times$10$^{11}$cm$^{-2}$ and $\mu$$\approx$2.6$\times$10$^5$cm$^2$/Vs.  It is from the same wafer as used in the experiments of Ref.~\cite{kang}, which found a RIP at $\nu$$>$1/3.

\begin{figure}[th]
\includegraphics[width=11.5cm]{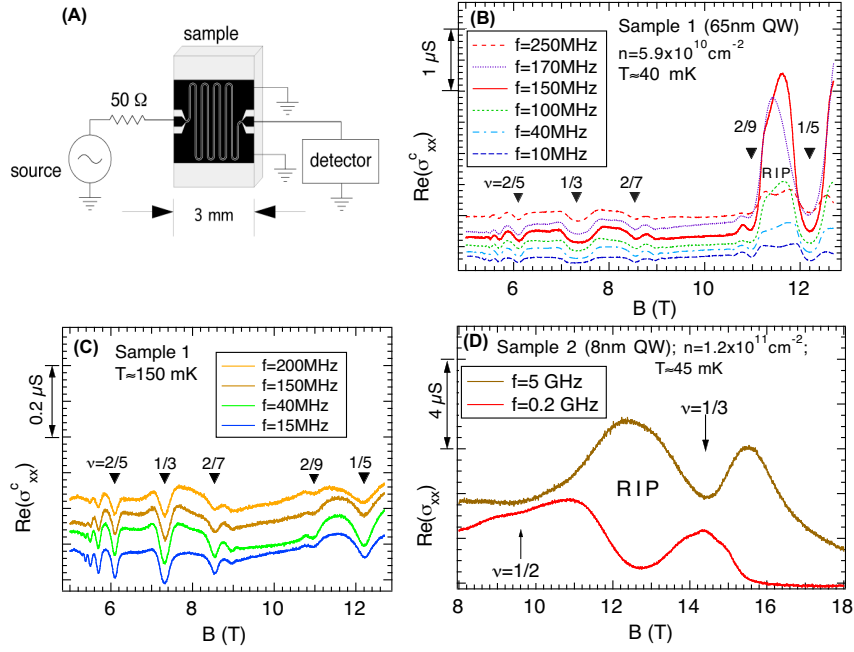}
\caption{\textbf{(A)}: Schematic of measurement circuit. Black regions represent the CPW metal films, consisting of the driven meander-shaped center line, separated from each of the two grounded side planes by a slot region (shown as the thin white meandering region) of width $w$. \textbf{(B)}: Sample 1 (65 nm QW), $B$-dependent Re($\sigma^c_{xx}$) measured at $T\approx$40 mK. Offset traces from bottom to top were measured at $f$=10, 40, 100, 150, 170 and 250 MHz respectively. The $f$=150 MHz trace shows a resonant enhancement of conductivity in the RIP, peaked at 
$B$$\sim$11.6T, $\nu$$\sim$0.21. \textbf{(C)}: Sample 1 measured at $T$$\approx$150 mK. The resonance seen at 40 mK nearly disappears. Offset traces from bottom to top were measured at $f$=10, 40, 150 and 200 MHz respectively. Note the difference in vertical scale with panel (B). 
\textbf{(D)}:Sample 2 (8 nm QW, with a RIP for 1/3$<$$\nu$$<$1/2). The lower trace shows magnetoconductivity at $f$=0.2 GHz, in comparison to the upper trace (offset) taken at 5 GHz. Measurements were done at $T\approx$45 mK.} \label{fig:fig1} 
\end{figure}

Figure~\ref{fig:fig1}(A) shows the scheme of our contactless, high frequency ($f$) (in RF/microwave range) magnetotransport measurements, employing similar techniques to those used previously.\cite{lloyd,wc} A meandering metal film coplanar waveguide (CPW) is deposited on the sample surface. A network analyzer generates an AC signal which propagates through 
the CPW, setting up an AC electric field mainly confined to the slots between the center conductor and the broad side planes.
The relative power absorption ($P$) by the 2DES is measured. 
Under conditions\cite{lloyd} of (1) high $f$, (2) low 2DES conductivity, (3) no reflections at ends of CPW, and (4) 2DES is in its long wavelength limit, one has $P=\mathrm{exp}(\frac{2lZ_0}{w}\mathrm{Re}(\sigma_{xx}))$ where Re($\sigma_{xx}$) is real part of the diagonal conductivity of the 2DES, $l$ and $w$ the total length (28mm) and slot width (30 $\mu$m)
of the CPW and $Z_0$=50 $\Omega$ its characteristic impedance at $\sigma_{xx}=0$. It turns out that conditions (1)-(4) are well satisfied for sample 2, allowing us to directly extract Re($\sigma_{xx}$) from $P$. The conditions are not fully satisfied\cite{wsa} for sample 1, nonetheless we cast the measured $P$ into a Re($\sigma^{c}_{xx}$)=$(w/2lZ_0)$ln($P$), and still refer to Re($\sigma^{c}_{xx}$) as ``conductivity'', which can actually differ from the true 2DES Re($\sigma_{xx}$) by a factor of order unity\cite{note}. In this work we focus on magnetoconductivity measurements (sweeping $B$ at different $f$'s), which are
complementary to spectroscopy measurements (sweeping $f$ at different $B$'s)\cite{wc,wsa}, so as to facilitate the comparison with DC transport. Measurements are performed close to the low power limit, by reducing RF/microwave power till $P$ no longer shows appreciable change. 
 
Fig.~\ref{fig:fig1}(B) shows magnetotransport traces measured on sample 1, 
at $T\approx$40 mK and several $f$'s ranging from 10 MHz to 250 MHz. The traces are vertically offset for clarity, and displayed from bottom to top in increasing order of $f$. Several representative filling factors are labeled, showing clearly resolved FQHE states at 1/3, 2/5, 2/7, 1/5 and 2/9, attesting to the high quality of the sample. For $\nu$$>$2/9 the conductivity (Re($\sigma^c_{xx}$)) shows only weak $f$-dependence. However, for $\nu$ between 2/9 and 1/5, in the RIP, large and nonmonotonic $f$-dependence is evident. Focusing on this region, we notice that at $f$=10 MHz, the conductivity is small, displaying a minimum near the center of the RIP. This resembles the 
behavior of low-$T$ DC conductivity.\cite{goldman} 
At higher $f$, such as 150 MHz, the conductivity is greatly enhanced, and displays a \textit{peak} near the center of the RIP ($B$$\sim$11.6 T with $\nu$$\sim$0.21). At even higher $f$, for example 250 MHz, the conductivity falls back to smaller values and displays again a minimum in the RIP.  The data thus reveal a clear resonance in Re($\sigma^c_{xx}$) (or absorption) near $f$$\sim $150 MHz. Similar resonant behavior is also noticed for $\nu$ below the 1/5 FQHE. The $f$ dependence for $\nu$ much lower than 1/5 has also been recently investigated\cite{wsa} and shows significantly different behavior, which is beyond the scope of this work. 

Fig.~\ref{fig:fig1}(C) shows magnetotransport at several $f$'s measured at elevated $T$$\approx$150 mK ($T$ accurate within 15\%).  The strong resonant behavior of RIP conductivity (near 150 MHz) observed at 40 mK is no longer evident.  

The observed resonant enhancement of low $T$ magnetoconductivity in the RIP near a well defined frequency supports the interpretation of such RIP as some disorder-pinned electron solid, with some order related to WC\cite{cfwc}. The resonance is interpreted as due to the pinning mode\cite{fuleenormand} of this solid. Probably because of lower disorder in sample 1, such a clear resonance in RIP was not observed in previous RF or microwave spectroscopy experiments\cite{ypli,wc} that were done on other wafers.  The observed (at temperature $T_\mathrm{exp}$$\sim$ 40-50 mK) resonant frequency $f$$\sim$150 MHz satisfies $hf\ll k_\mathrm{B}T_\mathrm{exp}$, which reflects the highly collective nature of such state and rules out individual particle localization-delocalization transition\cite{klz} giving the resonance. 


We have also investigated the high $f$ magnetotransport of sample 2, the narrow QW with
a RIP\cite{kang} above $\nu$=1/3. We find that strong $f$-dependence of magnetoconductivity is 
observed at much higher $f$ in sample 2 compared to sample 1. Fig.~\ref{fig:fig1}(D) shows two representative 
Re($\sigma_{xx}$) traces taken at $f$=0.2 GHz and 5 GHz. At $f$=0.2 GHz, the magnetoconductivity 
 displays a minimum in the RIP, at $\nu$$\sim$0.38 ($B$$\sim$12.6 T), the same $\nu$ 
at which experiments in Ref.~\cite{kang} found a peak in DC resistivity.  Interestingly, the magnetoconductivity displays a \textit{maximum} at $\nu$=1/3 instead of the expected FQHE minimum, which is recovered at elevated $T$. In contrast, at $f$=5 GHz, the magnetoconductivity in the RIP appears to be significantly enhanced and displays a peak, although we have not observed a clear resonant behavior up to 5 GHz.  A minimum at  $\nu$=1/3 is evident in the $f$=5 GHz trace. Hence
the abnormal behavior near $\nu$=1/3 may indicate that there is some nontrivial $f$-dependence even at  $\nu$=1/3, where DC transport in Ref.~\cite{kang} observed a  FQHE\cite{13note} . It is also possible that due to different cooldown and illumination procedures, the sample behaves differently in our experiments than that in Ref.~\cite{kang}. More work is clearly needed, especially extending the measurements to higher $f$ range and performing spectroscopy surveys, to better understand the RIP in the narrowly confined 2DES, and its likely intriguing relation with the $\nu$=1/3 FQHE.

\section*{Acknowledgements}
The high frequency measurements were performed at the National High Magnetic Field Lab, supported by NSF Cooperative Agreement No. DMR-0084173 and by State of Florida. We thank G.~Jones, T.~Murphy and E.~Palm for assistance and W.~Kang for helpful discussions. Financial support of this work was provided by AFOSR, DOE and NHMFL-IHRP.


\begin{thebibliography}{0}
\bibitem{note} This gives an uncertainty in the absolute magnitude of the conductivity data we present for sample 1, but has no significant influence on our findings regarding the $B$, $f$ and $T$ dependence of the conductivity.
\end{thebibliography}
\end{document}